\documentclass[twocolumn,preprintnumbers,showpacs,showkeys,amsmath,amssymb,superscriptaddress]{revtex4}
\usepackage{graphicx,dcolumn,bm,float}
\begin{document}
\title{Particle current fluctuations in a variant of asymmetric Glauber model}
\author{S. R. Masharian}
\email{masharian@iauh.ac.ir}
\affiliation{Islamic Azad University, Hamedan Branch, Hamedan, Iran}
\author{P. Torkaman}
\email{p.torkaman@basu.ac.ir}
\affiliation{Physics Department, Bu-Ali Sina University, 65174-4161 Hamedan, Iran}
\author{F. H. Jafarpour}
\email{farhad@ipm.ir}
\affiliation{Physics Department, Bu-Ali Sina University, 65174-4161 Hamedan, Iran}
\begin{abstract}
We study the total particle current fluctuations in a one-dimensional stochastic system of classical particles 
consisting of branching and death processes which is a variant of asymmetric zero-temperature Glauber 
dynamics. The full spectrum of a modified Hamiltonian, whose minimum eigenvalue generates the large deviation 
function for the total particle current fluctuations through a Legendre-Fenchel transformation, is obtained analytically. 
Three examples are presented and numerically exact results are compared to our analytical calculations. 
\end{abstract}
\maketitle
\section{Introduction}
Most of one-dimensional non-equilibrium systems with stochastic dynamics show unique collective behaviors in their steady-states which usually 
can not be found in their equilibrium counterparts. Non-equilibrium phase transition and shock formation are two examples of these remarkable 
behaviors. Non-equilibrium systems are also interesting to study from a mathematical point of view. These systems have opened up new horizon 
of research in the field of exactly solvable systems~\cite{P97,S01}.

During the last decade several non-equilibrium exactly solvable systems have been introduced and studied in related literature. On the other hand,
different mathematical techniques have been developed to study their steady-state properties. A matrix method,
known as the matrix product method, is introduced and used to calculate the steady-state and the average value of the physical quantities in the 
steady-state of these systems~\cite{BE07}. The excitations, which give the relaxation times, can also be obtained using the {\it Bethe Ansatz}~\cite{MG}. 
Recently, there has been attempts to establish connections between Bethe Ansatz and matrix  product method~\cite{GM}.
Other interesting quantities include the large deviation function for the probability distribution of fluctuating quantities, such as the particle current, in the 
steady-state of these systems~\cite{T09,BD07}. The large deviation function can be obtained, through a Legendre-Fenchel transformation, from 
the minimum eigenvalue of a modified Hamiltonian. We are then basically left with finding the minimum eigenvalue of a matrix. The number of 
systems for which this quantity can be calculated exactly is very limited.

In this paper we consider a stochastic system of classical particles in which the particles interact with each other according to a 
variant of the zero-temperature Glauber dynamics on a lattice with open boundaries~\cite{G63,KA01,KJS}. 
More precisely the particles are injected and extracted from the first and the last sites of the lattice respectively. In the bulk of the lattice, on the other 
hand, the particles are subjected to totally asymmetric branching and death processes. The steady-state of this system 
has already been calculated using the matrix product method~\cite{J04}. It is known that this steady-state can be written as a 
linear combination of product shock measures with one shock front and that these shock fronts have simple random walk dynamics~\cite{JM07}. 

We are specially interested in the total particle current fluctuations in this system. As we mentioned, the large deviation function for the 
particle current fluctuations can be obtained from the Legendre-Fenchel transformation of the minimum eigenvalue of a modified Hamiltonian.
This modified Hamiltonian can be constructed from the stochastic time evolution operator of the system (sometimes called the Hamiltonian). This can be 
done by multiplying the non-diagonal elements of the Hamiltonian of the system by an exponential factor which counts the 
particle jumps contributing to the total particle current, see for example~\cite{HS07,DL98,LS99}. It turns out that the modified Hamiltonian 
associated with the total particle current fluctuations can be fully diagonalized. The key point is to change the basis of the vector space in an 
appropriate way by introducing a product shock measure with multiple shock fronts. In this new basis the modified Hamiltonian becomes 
an upper block-bidiagonal matrix which is much easier to work with, because we only need to diagonalize the diagonal blocks.

Our analytical investigations reveal that for the small particle current fluctuations (smaller than the average particle current in the 
steady-state to be more precise) the eigenvector associated with the minimum eigenvalue of the modified Hamiltonian should 
be written as a linear combination of product shock measures with a single shock front. In contrast, for the large particle current 
fluctuations  (larger than the average particle current in the steady-state to be more precise) it should be written as a linear combination 
of product shock measures with more than one shock front. The validity of our analytical calculations is checked by comparing the analytical results 
with those obtained from numerical diagonalization of the modified Hamiltonian.

This paper is organized as follows. In the second section we will review the known results on the steady-state properties of the 
system. The total particle current is also introduced and its average value in the steady-state is calculated.  In the 
third section we will briefly review the basics of the particle current fluctuations. The forth section is devoted to the diagonalization 
of the modified Hamiltonian. The minimum eigenvalue of the modified Hamiltonian will be discussed in the fifth section. 
We will compare the analytical and numerical results in the sixth section. The concluding remarks are also given in the last section.

\section{Steady-State}
Let's consider a lattice of length $L$. We assume that each lattice site can be occupied by at most one particle or a vacancy. 
The reaction rules between two consecutive sites $k$ and $k+1$ on the lattice are as follow
\begin{equation}
\label{rules}
\begin{array}{ll}
A \; \emptyset \; \longrightarrow \; \emptyset \; \emptyset \quad \mbox{with the rate} \quad \omega_1 \\
A \; \emptyset \; \longrightarrow \; A \; A \quad \mbox{with the rate} \quad \omega_2.
\end{array}
\end{equation}
in which a particle (vacancy) is labeled with $A$ ($\emptyset$). A particle can enter the system from the left 
boundary of the lattice with the rate $\alpha$. A particle can also leave the system from the right boundary 
with the rate $\beta$. This model is an asymmetric variant of zero-temperature Glauber dynamics~\cite{G63,KA01,KJS}.
The time evolution of the probability distribution vector $\vert P(t) \rangle$ is given by a master equation~\cite{S01}
\begin{equation}
\label{master equation}
\frac{d}{dt}\vert P(t) \rangle=-H \vert P(t) \rangle
\end{equation}
where the Hamiltonian $H$ is an stochastic square matrix
\begin{equation}
\begin{array}{lll}
\label{hamiltonian}
H & = & \sum_{k=1}^{L-1} \big ( {\cal I}^{\otimes (k-1)}\otimes h\otimes {\cal I}^{\otimes (L-k-1)}\big )\\ \\
   & + & h^{(1)} \otimes {\cal I}^{\otimes (L-1)} \\ \\
   & + & {\cal I}^{\otimes (L-1)}\otimes h^{(L)}
\end{array}
\end{equation}
in which $\cal I$ is a $2 \times 2$ identity matrix and that we have defined 
$$
V^{\otimes N} \equiv \underbrace{V \otimes V\cdots \otimes V}_\text{$N$ times}. 
$$ 
Introducing the basis kets
$$
\vert  \emptyset \rangle = \left( \begin{array}{c}
1\\
0\\
\end{array} \right)\, ,\;\;  
\vert A \rangle=\left( \begin{array}{c}
0\\
1\\
\end{array} \right)\,
$$
the matrix representation of $h$ in the basis of $\{ \emptyset \emptyset, \emptyset A, A \emptyset, A A\}$ 
and that of $h^{(1)}$ and $h^{(L)}$ in the basis of $\{ \emptyset,A \}$ are given by
$$
\begin{array}{l}
h=\left( \begin{array}{cccc}
0 & 0&-\omega_{1}&0\\
0 & 0&0&0\\
0&0&\omega_{1}+\omega_{2}&0\\
0&0&-\omega_{2}&0\\
\end{array} \right)\, , \\ \\
h^{(1)}=\left( \begin{array}{cc}
\alpha & 0\\
-\alpha & 0\\
\end{array} \right)\, , \\ \\
h^{(L)}=\left( \begin{array}{cc}
0 & -\beta\\
0 & \beta\\
\end{array} \right)\, . 
\end{array}
$$
The right eigenvector with vanishing eigenvalue of Hamiltonian $H$ gives the steady-state probability distribution vector of the system.
It is  known that this vector can be written as a linear combination of product shock measures with 
a single shock front~\cite{JM07}. It turns out that the dynamics of the position of a shock front is similar to that of a biased 
random walker moving on a finite lattice with reflecting boundaries. 
This steady-state probability distribution vector has also been obtained using a matrix product method in~\cite{J04}. 
By associating the operators $E$ and $D$ to the presence of a vacancy and a particle in a given lattice site,  
the steady-state weight of any configuration $\{ \tau_1,\cdots,\tau_L \}$ is proportional to
\begin{equation}
\label{weight}
\langle\langle W \vert \prod_{k=1}^{L}(\tau_{k} D + (1-\tau_{k}) E) \vert V \rangle\rangle
\end{equation}
in which $\tau_{k}=0$ if the lattice-site $k$ is empty and $\tau_{k}=1$ if it is occupied by a particle. 
In (\ref{weight}), $\vert V \rangle\rangle$ and $\langle \langle W \vert$ are two auxiliary vectors. It has been shown
that these two operators and vectors have a two-dimensional matrix representation given by~\cite{J04}
\begin{equation}
\begin{array}{ll}
\label{representation}
D=\left( \begin{array}{cc}
0 & 0\\
d & \frac{\omega_{2}}{\omega_{1}}\\
\end{array} \right),\;\;
E=\left( \begin{array}{cc}
1 & 0\\
-d & 0\\
\end{array} \right),\\ \\
\vert V \rangle\rangle=\left( \begin{array}{cc} \frac{-\beta \omega_{2}}{(\omega_{2}-\omega_{1}+ \beta) d\omega_{1}}\\ 
1 \end{array} \right),\\ \\
\langle\langle W \vert=\left( \begin{array}{cc} \frac{(\omega_{1}-\omega_{2}+ \alpha)d}{\alpha} & 1 \end{array} \right)
\end{array}
\end{equation}
in which $d$ is a free parameter. Using~(\ref{weight}) and~(\ref{representation}) one can easily calculate the weight of any 
configuration in the steady-state and also the average value of the physical quantities, such as the particle current, in the long-time limit.

Let us call $\langle \rho_{k} \rangle(t)$ the average local density of particles at the lattice site $k$ at time $t$. Using~(\ref{rules}) and
considering the injection and extraction of particles at the boundaries the time evolution
of this quantity is given by
\begin{eqnarray}
\label{density evol}
\frac{d}{dt} \langle \rho_{1} \rangle(t)&=& \alpha  \langle 1-\rho_{1} \rangle-\omega_1 \langle \rho_{1}  (1-\rho_{2})\rangle, \nonumber\\
\frac{d}{dt} \langle \rho_{k} \rangle(t)&=& \omega_2  \langle \rho_{k-1}  (1-\rho_{k})\rangle-\omega_1 \langle \rho_{k}  (1-\rho_{k+1})\rangle, \\
\frac{d}{dt} \langle \rho_{L} \rangle(t)&=&\omega_2  \langle \rho_{L-1}  (1-\rho_{L})\rangle-\beta  \langle \rho_{L}  \rangle. \nonumber
\end{eqnarray}
in which $k=2,\cdots,L-1$. The average local density of particles is related to the average particle current through a continuity equation
\begin{equation}
\label{current-density}
\frac{d}{dt} \langle \rho_{k} \rangle(t)=  \langle J_{k-1} \rangle(t)- \langle J_{k} \rangle(t)+S_{k}(t) 
\end{equation}
for $k=1,\cdots,L$. We define $\langle J_{k} \rangle(t)$ as the average local particle current from the lattice site $k$ to $k+1$ at time $t$. 
$S_{k}(t)$ is also called a source term. In~(\ref{current-density}) we have also assumed that $\langle J_{0} \rangle(t)=\langle J_{L} \rangle(t)=0$. 
In the steady-state the time dependency of the quantities will be dropped and we find
\begin{equation}
\label{source1}
S_{k}=  \langle J_{k} \rangle- \langle J_{k-1} \rangle \,\, \mbox{for} \,\, k=1,\cdots,L \, .
\end{equation}
Comparing~(\ref{density evol}) and~(\ref{current-density}) one finds the following relation for the average particle current in the steady-state 
\begin{equation}
\label{current}
\langle J_{k} \rangle= (\omega_{1}+\omega_{2}) \langle \rho_{k}(1-\rho_{k+1}) \rangle \,\, \mbox{for} \,\, k=1,\cdots,L-1 
\end{equation}
and also the source terms which are defined as follows
\begin{eqnarray}
S_{1}&=& \alpha \langle 1-\rho_{1} \rangle +\omega_{2} \langle \rho_{1}(1-\rho_{2})\rangle, \nonumber \\
S_{k}&=& \omega_{2}   \langle \rho_{k}(1-\rho_{k+1}) \rangle - \omega_{1}  \langle \rho_{k-1}(1-\rho_{k}) \rangle,\\
S_{L}&=& -\beta \langle \rho_{L} \rangle -\omega_{1} \langle \rho_{L-1} (1-\rho_{L}) \rangle. \nonumber 
\end{eqnarray}
in which $k=2,\cdots,L-1$. The average particle current in~(\ref{current}) can be understood by investigating~(\ref{rules}). 
The second dynamical rule in~(\ref{rules}) clearly increase the particle current. The first dynamical rule in~(\ref{rules}) can 
be considered as a backward movement of a vacancy. This is equivalent to a forward movement of a particle which, again, 
increases the particle current. Similar examples of particle current in the presence of source terms can be found in~\cite{SA}.

The average local particle current in the steady-state can be calculated using the matrix product method. The result is
\begin{widetext}
\begin{eqnarray*}
\langle J_{k} \rangle 
& = & (\omega_{1}+\omega_{2}) \langle \rho_{k}(1-\rho_{k+1}) \rangle  \\
& = & (\omega_{1}+\omega_{2}) \frac{\sum_{\{ \tau \}} \langle \langle W \vert \prod_{i=1}^{k-1} (\tau_{i} D + (1-\tau_{i}) E) 
DE \prod_{i=k+2}^{L} (\tau_{i} D + (1-\tau_{i}) E)  \vert V \rangle \rangle }
{\sum_{\{ \tau \}} \langle \langle W \vert \prod_{i=1}^{L} (\tau_{i} D + (1-\tau_{i}) E)  \vert V \rangle \rangle}\\
& = &(\omega_1+\omega_2)\frac{\langle\langle W\vert C^{k-1}DEC^{L-k-1} \vert V\rangle\rangle}{\langle\langle W\vert C^{L}\vert V \rangle\rangle}\\
& = &\frac{(\omega_{2}^{2}-\omega_{1}^{2}) \alpha \beta(\frac{\omega_2}{\omega_1})^{k}}
{\alpha \omega_1(\beta-\omega_1+\omega_2)(\frac{\omega_2}{\omega_1})^{L}-\beta \omega_2 (\alpha+\omega_1-\omega_2)} 
\end{eqnarray*}
\end{widetext}
in which we have defined $C=D+E$. Defining the total particle current as
\begin{equation}
\label{total current}
\langle J \rangle=\sum_{k=1}^{L-1}\langle J_{k} \rangle
\end{equation}
it is easy to see that in the limit of $L\rightarrow \infty$ we have
\begin{equation}
\label{average current}
\langle J \rangle=\left\{
\begin{array}{ll}
\frac{\beta(\omega_1 +\omega_2)}{\beta-\omega_1+\omega_2} & \quad \mbox{for} \quad \omega_1 < \omega_2 \, , \\  \\
2\omega_{1} & \quad \mbox{for} \quad \omega_{1}=\omega_{2}\, , \\ \\
\frac{\alpha(\omega_1 +\omega_2)}{\alpha+\omega_1-\omega_2} & \quad \mbox{for}\quad \omega_1 > \omega_2 \, .
\end{array}
\right.
\end{equation}
This indicates that the system undergoes a phase transition at $\omega_{1}=\omega_{2}$. The phase $\omega_{1} > \omega_{2}$ 
($\omega_{1} < \omega_{2}$) is called the low-density (high-density) phase. 
\begin{figure*}
\begin{centering}
\includegraphics[width=5in]{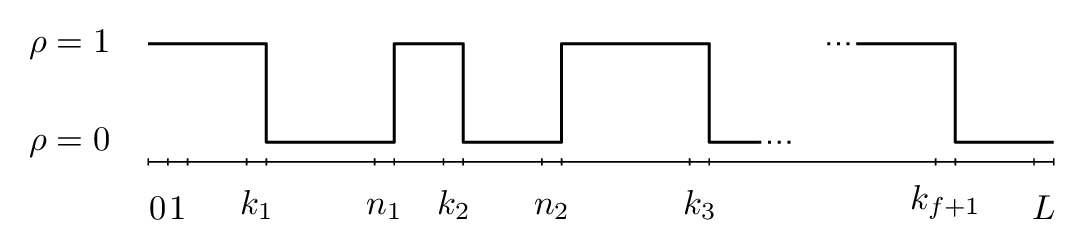}
\caption{\label{fig1} A simple sketch of a base vector defined in~(\ref{basis}).}
\end{centering}
\end{figure*}
%
\section{Particle Current Fluctuations}
Assuming that the large deviation principle holds, the probability distribution for observing the total particle current $J$ in the system is given by
\begin{equation}
P(J) \approx e^{- t\, I(J)}
\end{equation}
which is valid for $ t \to \infty$. The large deviation function $I(J)$ measures the rate at which the total particle current deviates from its average value.
It is known that the large deviation function $I(J)$ is, according to the G{\"a}rtner-Ellis theorem, the Legendre-Fenchel transform of the minimum
eigenvalue of a modified Hamiltonian $\tilde{H}$, denoted by $\Lambda^{\ast}(\lambda)$,~\cite{T09} 
\begin{equation}
\label{GE}
I(J)=\max_{\lambda}(\Lambda^{\ast}(\lambda)-J \lambda) \; .
\end{equation}
The modified Hamiltonian is defined as follows
\begin{equation}
\label{modified hamiltonian}
\begin{array}{lll}
{\tilde H} & = & \sum_{k=1}^{L-1} \big ( {\cal I}^{\otimes (k-1)}\otimes {\tilde h}\otimes {\cal I}^{\otimes (L-k-1)}\big ) \\ \\
               & + & {\tilde h}^{(1)} \otimes {\cal I}^{\otimes (L-1)} \\ \\
               & + & {\cal I}^{\otimes (L-1)}\otimes {\tilde h}^{(L)}
\end{array}
\end{equation}
in which 
$$
\begin{array}{l}
{\tilde h}=\left( \begin{array}{cccc}
0 & 0&-\omega_{1}e^{-\lambda}&0\\
0 & 0&0&0\\
0&0&\omega_{1}+\omega_{2}&0\\
0&0&-\omega_{2}e^{-\lambda}&0\\
\end{array} \right)\, , \\ \\
{\tilde h}^{(1)}=\left( \begin{array}{cc}
\alpha & 0\\
-\alpha & 0\\
\end{array} \right)\, , \\ \\
{\tilde h}^{(L)}=\left( \begin{array}{cc}
0 & -\beta\\
0 & \beta\\
\end{array} \right)\, . 
\end{array}
$$
Note that since the modified Hamiltonian defined in~(\ref{modified hamiltonian}) becomes equal to the 
stochastic time evolution operator given in~(\ref{hamiltonian}) at $\lambda=0$ then we have 
$\Lambda^{\ast}(\lambda=0)=0$. Finally, the first derivative of the minimum eigenvalue 
respect to $\lambda$ at $\lambda=0$ gives the total current defined in~(\ref{total current})                                                  
\begin{equation}
\label{average J}
\langle J \rangle=\frac{d\Lambda^{\ast}}{d \lambda} \Big\vert_{\lambda=0}.
\end{equation}
In the following section we will show that $\tilde{H}$, defined in~(\ref{modified hamiltonian}), 
can be diagonalized exactly.

\section{Diagonalization of $\tilde{H}$}
In order to diagonalize the modified Hamiltonian $\tilde{H}$ defined in~(\ref{modified hamiltonian}) we start with redefining the basis of the vector 
space by introducing the following product shock measure with $N=2f+1$ shock fronts
\begin{widetext}
\begin{equation}
\label{basis}
\vert k_{1},n_{1},k_{2},n_{2},\cdots,n_{f},k_{f+1} \rangle=
\vert A \rangle^{\otimes k_{1}} 
\otimes 
\vert \emptyset \rangle^{\otimes (n_{1}-k_{1})}
\otimes 
\vert A \rangle^{\otimes (k_{2}-n_{1})}
\otimes 
\cdots
\otimes 
\vert A \rangle^{\otimes (k_{f+1}-n_{f})}
\otimes 
\vert \emptyset \rangle^{\otimes (L-k_{f+1})}
\end{equation}
\end{widetext}
in which $N=1,3,5,\cdots,L+1$ for an even $L$ and $N=1,3,5,\cdots,L$ for an odd $L$ and that 
$0 \le k_{1} < n_{1} < k_{2} < \cdots<n_{f} < k_{f+1} \le L$.  A simple sketch of such a product 
shock measure with multiple shock fronts is given in Fig.~\ref{fig1}.

For a given $N$ the number of these vectors is simply given by a Binomial coefficient $C_{L+1,N}\equiv(L+1)!/(N!(L+1-N)!)$. 
Now the dimensionality of the vector space constructed with these vectors can be obtained as
\begin{equation}
\sum_{N}C_{L+1,N}=2^L
\end{equation}
regardless of whether $L$ is even or odd. The vectors~(\ref{basis}) make a complete 
orthonormal basis for our $2^L$-dimensional vector space. Assuming $L$ is an even number~\footnote{Our approach is 
not affected when the system size is an odd number. One should only consider $N=1,3,\cdots,L$ everywhere throughout 
this section.}, in the basis 
\begin{equation}
\{ \vert k_{1} \rangle, \vert k_{1},n_{1},k_{2}\rangle, \cdots, \vert k_{1}, n_{1},\cdots,n_{f},k_{f+1} \rangle \}
\end{equation}
the modified Hamiltonian~(\ref{modified hamiltonian}) has the following upper block-bidiagonal matrix representation
\begin{equation}
\label{new modified hamiltonian}
\tilde{H}=
\left( \begin{array}{ccccccc}
A_{1} & B_{3}&0&0&\cdots&0&0\\
0 & A_{3}&B_{5}&0&\cdots&0&0\\
0&0&A_{5}&B_{7}&\cdots&0&0\\
\vdots&&&\ddots&\ddots&& \vdots\\
0&0&0&0&\cdots&B_{L-1}&0\\
0&0&0&0&\cdots&A_{L-1}&B_{L+1}\\
0&0&0&0&\cdots&0&A_{L+1}
\end{array} \right)\,
\end{equation}
in which $A_{N}$ for $N=1,3,\cdots,L+1$ is a $C_{L+1,N}\times C_{L+1,N}$ matrix and $B_{N}$ for $N=3,5,\cdots,L+1$
is a $C_{L+1,N-2}\times C_{L+1,N}$ matrix. The matrix elements of these matrices are given explicitly in the Appendix~\ref{A1}. 

Although the modified Hamiltonian~(\ref{new modified hamiltonian}) is not a stochastic matrix; however,
its matrix structure suggests the following picture which will be discussed in more detail in the forthcoming sections. 
Acting~(\ref{new modified hamiltonian}) on $\vert k_{1} \rangle$ with $k_1=0,1,\cdots,L$, which is a product 
shock measure with a single shock front, gives a linear combination of product shock measures with a single 
shock front. These series of evolution equations are quite similar to the evolution equations for a particle at the 
lattice site $k_{1}$ performing a biased random walk on a one-dimensional lattice with reflecting boundaries. 
In other words, the vectors $\{ \vert k_{1} \rangle \}$ define an invariant sector, which will be called 
${\cal S}_1$, in the sense that acting $\tilde{H}$ on any member of this sector gives a linear combination of the vectors 
in the same sector. The matrix elements of $A_{1}$ in~(\ref{new modified hamiltonian}) determine the coefficients of these 
linear expansions. 

On the other hand, acting~(\ref{new modified hamiltonian}) on $\vert k_{1},n_{1},k_{2} \rangle$ 
with $0 \le k_1 < n_1 < k_2 \le L$, which is a product shock measure with three shock fronts, gives
a linear combination of the product shock measures with a single or three shock fronts. 
These series of evolution equations are quite similar to those of two random walkers at the lattice sites 
$k_{1}$ and $k_{2}$ besides an obstacle at the lattice site $n_{1}$ which does not have any dynamics. 
The reason that the shock front at the lattice site $n_{1}$ (or the obstacle) does not have any dynamics can be easily 
understood by looking at~(\ref{rules}). In fact, the position of a shock front of type $0\cdots01\cdots1$ is 
not affected by these dynamical rules. 
As long as the random walkers are more than a single lattice site away from the obstacle, they perform biased random 
walks on the lattice. The random walkers also reflect from the boundaries of the lattice whenever they 
reach to the boundaries. When one of the random walkers arrives at an obstacle, the random walker  
and the obstacle both disappear; however, the other random walker continues to perform a biased 
random walk on the lattice. No new random walker or obstacle will be created once they disappear. 
The matrices $A_{3}$ and $B_{3}$ are responsible for the dynamics of these random walkers.
The above argument suggests that the vectors in ${\cal S}_1$ besides the vectors $\{ \vert k_{1},n_{1},k_{2} \rangle \}$ define an 
invariant sector, which will be called ${\cal S}_3$. Acting $\tilde{H}$ on any member of this sector gives a linear combination 
of the vectors in the same sector.

The next invariant sector, which will be called ${\cal S}_5$, is defined by the vectors in ${\cal S}_3$
and $\{ \vert k_{1},n_{1},k_{2},n_{2},k_{3} \rangle \}$ in which $0 \le k_1 < n_1 < k_2 < n_2 < k_3 \le L $. 
In this case we have three random walkers which are separated from each other by two obstacles. Once
a random walker meets an obstacle at two consecutive lattice sites, they disappear. The matrices $A_{5}$
and $B_{5}$ generates the dynamics of the random walkers and their interactions with the obstacles. 

This procedure can be continued to see that there are $L/2+1$ invariant sectors. 

In order to find the eigenvalues of the modified Hamiltonian~(\ref{new modified hamiltonian}) we can follow 
two equivalent scenarios. From one hand, the eigenvectors of this matrix can be written as a linear combination 
of the vectors in each invariant sector. This helps us find all of its eigenvalues. 
On the other hand, since the eigenvalues of the modified Hamiltonian are equal to those of $A_{N}$'s for $N=1,3,\cdots,L+1$, one 
can diagonalize each $A_{N}$ separately to calculate the eigenvalues of~(\ref{new modified hamiltonian}). We will employ the 
second approach which is the subject of the forthcoming sections.

\subsection{Diagonalization of $A_{1}$}
$A_{1}$ is a $C_{L+1,1} \times C_{L+1,1}$ tridiagonal matrix which has, in the basis~(\ref{basis}), the following matrix representation 
$$
A_{1}=\left( \begin{array}{ccccccc}
\alpha &-\omega_{1}e^{-\lambda}&0&\cdots&0&0&0\\
-\alpha &  \omega_{1}+\omega_{2}&-\omega_{1}e^{-\lambda}&\cdots&0&0&0\\
0 &  -\omega_{2}e^{-\lambda}&\omega_{1}+\omega_{2}&\cdots&0&0&0\\
\vdots & \vdots&\vdots&\ddots&\vdots &\vdots &\vdots\\
0&0&0&\cdots&\omega_{1}+\omega_{2}&-\omega_{1}e^{-\lambda} &0\\
0&0&0&\cdots&-\omega_{2}e^{-\lambda}&\omega_{1}+\omega_{2}& -\beta\\
0&0&0&\cdots&0 & -\omega_{2}e^{-\lambda}&\beta
\end{array} \right)\,
$$
The structure of this matrix reminds us of the evolution operator for a biased random walk moving on a one-dimensional 
lattice of length $L+1$ with reflecting boundaries  \textendash although the reader should note that the evolution operator 
is not a stochastic matrix. In an appropriate basis $\{ \vert 0 \rangle , \vert 1 \rangle ,\cdots , \vert L \rangle \}$ 
the evolution equations for the position of the random walker can be formally written as follows
$$
\begin{array}{l}
A_{1}\vert 0 \rangle = -\alpha\vert 1 \rangle + \alpha \vert 0 \rangle,  \\
A_{1}\vert k_{1} \rangle = -\omega_{1} e^{-\lambda} \vert k_{1}-1 \rangle-\omega_{2} e^{-\lambda} \vert k_{1}+1 \rangle + 
(\omega_{1}+\omega_{2})\vert k_{1} \rangle,\\ 
A_{1}\vert L \rangle =-\beta \vert L-1 \rangle+\beta \vert L \rangle
\end{array}
$$
with $k_{1}=1,\cdots,L-1$. Now the eigenvectors and also the eigenvalues of $A_{1}$ can be obtained using the same 
approach employed in~\cite{AA06} by writing 
\begin{equation}
\label{eigenvalue eq}
A_{1} \vert \Lambda_{1} \rangle =\Lambda_{1}(\lambda) \vert \Lambda_{1} \rangle
\end{equation}
and considering
\begin{equation}
\label{superposition}
\vert \Lambda_{1} \rangle=\sum_{k_{1}=0}^L C_{k_{1}} \vert k_{1} \rangle \; .
\end{equation}
Substituting~(\ref{superposition}) in~(\ref{eigenvalue eq}) and using the evolution equations for the shock front one can calculate 
$C_{k_{1}}$'s by applying a plane wave Ansatz. Defining 
$$
\eta \equiv \sqrt{\frac{\omega_{2}}{\omega_{1}}},
\quad \zeta \equiv 1-\frac{\alpha}{\omega_{2}}e^{\lambda},
\quad \xi \equiv 1-\frac{\beta}{\omega_{1}}e^{\lambda}
$$
and 
$$
F(x,y,z) \equiv e^{-\lambda } \left(y +x+(x^{-1} -y) z  \right)-(y+y^{-1}) 
$$
we find 
\begin{equation}
\label{cs}
C_{k_{1}}=\eta^{k_{1}} \frac{a(z_{1}) z_{1}^{k_{1}}+a(z_{1}^{-1})z_{1}^{-k_{1}}}{(1-\zeta)^{\delta_{k_{1},0}}(1-\xi)^{\delta_{k_{1},L}}} 
\end{equation}
in which 
$$
\frac{a(z_{1})}{a(z_{1}^{-1})}=-\frac{F(z_{1},\eta,\zeta)}{F(z_{1}^{-1},\eta,\zeta)} =-z_{1}^{-2L}\frac{F(z_{1}^{-1},\eta^{-1},\xi)}{F(z_{1},\eta^{-1},\xi)}  \; .
$$
It also turns out that the eigenvalues of $A_{1}$ are given by
\begin{equation}
\label{eigenvalue A1}
\Lambda_1(\lambda)=(\omega_{1}+\omega_{2})-e^{-\lambda}\sqrt{\omega_{1} \omega_{2}}(z_{1}+z_{1}^{-1}).
\end{equation}
The equation governing $z_{1}$ is also given by
\begin{equation}
\label{eq for zs}
z_{1}^{2L}=\frac{F(z_{1}^{-1},\eta,\zeta)F(z_{1}^{-1},\eta^{-1},\xi)}{F(z_{1},\eta,\zeta)F(z_{1},\eta^{-1},\xi)} \; .
\end{equation}
It can be seen that the equation~(\ref{eq for zs}) has $2L+4$ solutions. Two of these solutions i.e. $z_{1}=\pm 1$ have to be excluded
since for these values of $z_{1}$ the corresponding eigenvector vanishes. On the other hand, if $z_{1}$ is solution for~(\ref{eq for zs})
then $z_{1}^{-1}$ is also a solution. This means that the remaining $2L+2$ solutions result in $L+1$ eigenvalues. For $\lambda=0$ the pair
$z_{1}=\eta^{\pm 1}$ corresponds to the eigenvalue $\Lambda_1=0$. Finally, it can be shown that the
solutions of the equation~(\ref{eq for zs}) are either phases i.e. $\vert z_{1} \vert=1$ or they are real numbers. 
For the phase solutions $z_{1}=e^{i \theta}$ and the smallest eigenvalue in this case is given by
\begin{equation}
\label{phase sol}
\Lambda_1^{\text{phase}}(\lambda)=(\omega_{1}+\omega_{2})-2e^{-\lambda}\sqrt{\omega_{1} \omega_{2}} \; .
\end{equation}
The real solutions of the equation~(\ref{eq for zs}) are much easier to be found in the thermodynamic limit $L \to \infty$. 
Let us restrict the real solutions to $\vert z_{1} \vert >1$. It turns out that the equation~(\ref{eq for zs}) has
two real solutions in the thermodynamic limit
\begin{equation}
\label{z1 z2}
z_{1}^{(1)}=G(\eta^{-1},\zeta),\;\;\; z_{1}^{(2)}=G(\eta,\xi)
\end {equation}
where $G(x,y)$ is defined as follows
$$
\begin{array}{lll}
G(x,y)& = & \frac{1}{2x} \Big(e^{\lambda}(1+x^{2})+(y-1)\\ \\
          &+ & \sqrt{(e^{\lambda}(1+x^{2})+(y-1))^2-4 x^{2} y }\Big).
\end{array}
$$
Substituting~(\ref{z1 z2}) in~(\ref{eigenvalue A1}) gives the corresponding eigenvalues which will be denoted by $\Lambda_{1}^{(1)}$ 
and $\Lambda_{1}^{(2)}$. Whenever these eigenvalues exist, they will be definitely smaller than 
$\Lambda_{1}^{\text{phase}}(\lambda)$. The conditions under which $\Lambda_{1}^{(1)}$ and $\Lambda_{1}^{(2)}$ exist, will be discussed later. 

\subsection{Diagonalization of $A_{3}$}
In the basis~(\ref{basis}) the matrix $A_{3}$ is a $C_{L+1,3}\times C_{L+1,3}$ block diagonal matrix with $L-1$ blocks. These blocks will be called  
$A_{3}^{(n_{1})}$ with $n_{1}=1,\cdots,L-1$. On the other hand, for a given $n_{1}$, $A_{3}^{(n_{1})}$ is a $n_{1}(L-n_{1})$-dimensional 
block tridiagonal matrix with the following structure
$$
A_{3}^{(n_{1})}=\left( \begin{array}{ccccccc}
\tilde{A}_{1} & \tilde{B}_{1}&0&\cdots&0&0&0\\
\tilde{B}_{0} &  \tilde{A}_{2}& \tilde{B}_{1}&\cdots&0&0&0\\
0&\tilde{B}_{2}&\tilde{A}_{2}&\cdots&0&0&0\\
\vdots& \vdots& \vdots&\ddots&\vdots&\vdots &\vdots\\
0&0&0&\cdots& \tilde{A}_{2}& \tilde{B}_{1}&0\\
0&0&0&\cdots&\tilde{B}_{2}& \tilde{A}_{2}& \tilde{B}_{1}\\
0&0&0&\cdots&0&\tilde{B}_{2}&\tilde{A}_{2}
\end{array} \right)\, 
$$
in which ${\tilde A}_{1}$ and ${\tilde A}_{2}$ are $(L-n_{1}) \times (L-n_{1})$ square matrices whose matrix 
representations are given in Appendix~\ref{A2}. On the other hand, we have defined
$\tilde{B}_0=-\alpha {\cal I}$, $\tilde{B}_1=-\omega_{1}e^{-\lambda}{\cal I}$ and 
$\tilde{B}_2=-\omega_2 e^{-\lambda}{\cal I}$ where ${\cal I}$ is a $(L-n_{1}) \times (L-n_{1})$ identity matrix. Noting that 
$$
\sum_{n_{1}=1}^{L-1} n_{1}(L-n_{1})=C_{L+1,3}
$$ 
it is easy to check the dimensionality of $A_{3}$. 

Investigating the structure of $A_{3}^{(n_{1})}$ for a given $n_{1}$ suggests that it can be regarded as a 
non-stochastic evolution operator for two biased random walkers, moving on a one-dimensional lattice of 
length $L + 1$ with reflecting boundaries, which are separated by an obstacle. Let us denote the position of the first 
and the second random walker on the lattice by $k_1$ and $k_2$ respectively. The obstacle is at the lattice site $n_{1}$. 
For a fixed $n_{1}$ the random walkers can only hop into the lattice sites which satisfy the condition $0 \le k_1 < n_{1} < k_2 \le L$.
In terms of the matrix elements of $A_{3}^{(n_{1})}$, both random walkers reflect from the boundaries and also the obstacle.
One should note that the matrix $A_{3}$, in contrast to $\tilde{H}$, does not allow the random walkers to merge with the obstacles.
We remind the reader that $B$'s in~(\ref{new modified hamiltonian}) were responsible for disappearance of the random walkers
and the obstacles.

For a given $n_{1}$ let us introduce an appropriate $n_{1}(L-n_{1})$-dimensional basis $\{ \vert k_{1},k_{2} \rangle \}$ with
$0 \le k_{1} \le n_{1}-1$ and $n_{1}+1 \le k_{2} \le L$. We arrange these vectors as 
$\{ \vert 0,n_{1}+1 \rangle,\vert 0, n_{1}+2\rangle, \cdots \vert 0 ,L\rangle,
     \vert 1,n_{1}+1 \rangle,\vert 1, n_{1}+2\rangle, \cdots \vert 1 ,L\rangle,\cdots,
     \vert n_{1}-1, n_{1}+1\rangle,\vert n_{1}-1, n_{1}+2 \rangle,\cdots,\vert n_{1}-1, L \rangle \}$.
In this basis the evolution equations for the random walkers can be written as follows
\begin{widetext}
\begin{eqnarray} 
\label{eq A3}
A_{3}^{(n_{1})} \vert k_{1}, k_{2} \rangle &=&-\omega_1 e^{-\lambda} \vert k_{1}-1,k_{2} \rangle -\omega_2 e^{-\lambda} (1-\delta_{k_{1},{n_{1}-1}})  
\vert k_{1}+1,k_{2}\rangle \nonumber \\&-& \omega_1e^{-\lambda} (1-\delta_{n_{1}+1,k_{2}})\vert k_{1},k_{2}-1\rangle -\omega_2 e^{-\lambda} 
\vert k_{1},k_{2}+1 \rangle \nonumber  \\ &+& 2(\omega_1+\omega_2) \vert k_{1},k_{2} \rangle \;\; \mbox{for} \;\; 1 \le k_{1} \le n_{1}-1 , n_{1}+1 
\le k_{2} \le L-1\; ,\nonumber \\
A_{3}^{(n_{1})} \vert 0, k_{2} \rangle &=&-\alpha (1-\delta_{n_{1},1})\vert 1,k_{2}\rangle 
-\omega_1 e^{-\lambda}(1-\delta_{n_{1}+1,k_{2}})\vert 0,k_{2}-1\rangle  \nonumber \\
&-&\omega_2 e^{-\lambda}\vert 0,k_{2}+1\rangle \nonumber
+(\alpha+\omega_1+\omega_2)\vert 0,k_{2} \rangle \; \mbox{for} \;\; n_{1}+1 \le k_{2} \le L-1 \;\; , \\ 
A_{3}^{(n_{1})} \vert k_{1}, L \rangle &=&-\beta (1- \delta_{n_{1},L-1})\vert k_{1},L-1\rangle 
-\omega_2e^{-\lambda} (1-\delta_{k_{1},n_{1}-1})\vert k_{1}+1,L\rangle \nonumber \\
&-&\omega_1e^{-\lambda} \vert k_{1}-1,L\rangle
+(\beta+\omega_1+\omega_2) \vert k_{1},L\rangle \;\; \mbox{for} \;\; 1 \le k_{1} \le n_{1}-1 \; ,\\
A_{3}^{(n_{1})} \vert 0,L \rangle &=& -\alpha(1-\delta_{n_{1},1})\vert 1,L\rangle
-\beta  (1-\delta_{n_{1},L-1})\vert 0,L-1\rangle + (\alpha+\beta) \vert 0,L \rangle \; . \nonumber
\end{eqnarray}
\end{widetext}
These equations can be used to find the eigenvalues and eigenvectors of $A_{3}^{(n_{1})}$. The
reader can easily convince himself that in the above mentioned basis, the matrices $\tilde{A}_{1,2}$'s 
are responsible for moving the position of the second random walker while $\tilde{B}_{0,1,2}$'s 
are responsible for moving the position of the first random walker. For a given $n_{1}$, the eigenvalue equation 
\begin{equation}
A_3^{(n_{1})} \vert \Lambda_{3} \rangle=\Lambda_{3}(\lambda) \vert \Lambda_{3} \rangle
\end{equation}
can now be solved by using~(\ref{eq A3}) and introducing
\begin{equation}
\vert \Lambda_3 \rangle= \sum_{k_{1}=0}^{n_{1}-1}\sum_{k_{2}=n_{1}+1}^{L} C_{k_{1},k_{2}} \vert k_{1},k_{2} \rangle \; .
\end{equation}
By considering a plane wave ansatz and after some straightforward calculations one finds 
that, for a given $n_{1}$, the coefficients $C_{k_{1},k_{2}}$  are given by
$$
C_{k_{1},k_{2}}=\frac{\prod_{i=1}^{2} \eta^{k_{i}} \Big( a_i(z_i)z_i^{k_{i}}+a_i(z_i^{-1}) z_i^{-k_{i}} \Big )}
{(1-\zeta)^{\delta_{k_{1},0}}(1-\xi)^{\delta_{k_{2},L}}}
$$
for $0 \le k_{1} \le n_{1}-1$ and $n_{1}+1 \le k_{2} \le L$ in which
$$
\begin{array}{l}
\frac{a_1(z_1)}{a_1(z_1^{-1})}=-\frac{F(z_1,\eta,\zeta)}{F(z_1^{-1},\eta,\zeta)}=-z_1^{-2n_{1}} \; ,  \\ \\
\frac{a_2(z_2)}{a_2(z_2^{-1})}=-z_2^{-2L}\frac{F(z_2^{-1},\eta^{-1},\xi)}{F(z_2,\eta^{-1},\xi)}=-z_2^{-2n_{1}} \; .
\end{array}
$$
The eigenvalues of $A_{3}^{(n_{1})}$ are also given by
\begin{equation}
\label{eigenvalue A3}
\Lambda_{3}(\lambda)=2(\omega_1+\omega_2)-e^{-\lambda}\sqrt{\omega_1 \omega_2}(z_1+z_{1}^{-1}+z_2+z_{2}^{-1}) 
\end{equation}
in which the equations governing $z_1$ and $z_2$ are
\begin{equation}
\label{z1z2 eq}
\begin{array}{l}
z_1^{2 n_{1}}= \frac{F(z_1^{-1},\eta,\zeta)}{F(z_1,\eta,\zeta)} \; , \\ \\
z_2^{2(L- n_{1})}= \frac{F(z_2^{-1},\eta^{-1},\xi)}{F(z_2,\eta^{-1},\xi)} \; .
\end{array}
\end{equation}
The first equation in~(\ref{z1z2 eq}) has $2n_{1}+2$ solutions while the second equation has $2(L-n_{1})+2$ solutions. 
Excluding the solutions $z_1=\pm 1$ and $z_2=\pm 1$ and noting that if the pair $(z_1,z_2)$ is a solution then the pairs 
$(z_1^{-1},z_2^{-1})$, $(z_1,z_2^{-1})$ and $(z_1^{-1},z_2)$ are also the solution, one finds $n_{1}(L-n_{1})$ solutions 
(or eigenvalues) by mixing the solutions of the equations~(\ref{z1z2 eq}). 

In summary, for each $n_{1}=1,\cdots,L-1$ one solves the equations~(\ref{z1z2 eq}) to find $z_{1}$ and $z_{2}$. Substituting 
these into~(\ref{eigenvalue A3}) gives the corresponding eigenvalues. The total number of eigenvalues of $A_{3}$ obtained in this way will be $L(L^2-1)/6$.

\subsection{Diagonalization of $A_{N}$}
In the basis~(\ref{basis}) the matrix $A_{N}$ is a $C_{L+1,N}\times C_{L+1,N}$ block diagonal matrix.  
Our procedure in the preceding sections can be continued to see that each block of $A_{N}$ for $N=1,3,\cdots,L+1$ can 
be regarded as a non-stochastic evolution operator for $f+1=(N+1)/2$ biased random walkers at the positions 
$\{ k \} =\{ k_{1}, k_{2},\cdots, k_{f+1} \}$ which are separated by $f$ obstacles at the positions $\{ n \}=\{ n_{1},n_{2},\cdots,n_{f} \}$ 
given that $0 \le k_{1} < n_{1} < k_{2} < n_{2} < \cdots < n_{f} < k_{f+1} \le L$. In other words, for $i=1,2,\cdots,f+1$ the positions
of the obstacles and the random walkers should satisfy the following constraints 
\begin{equation}
\label{position}
\begin{array}{l}
n_{i-1}+2 \le n_{i} \le L-(N-2i)  \; ,\\ \\
n_{i-1}+1 \le k_{i} \le n_{i}-1
\end{array}
\end{equation}
with $n_{0}\equiv -1$ and $n_{f}\equiv L+1$. For a given $N$ each block of $A_{N}$, which will be called $A_{N}^{(\{ n \})}$, is a 
$D_{\{ n \} }$-dimensional square matrix where $D_{\{ n \} }$ is given by
$$
D_{\{  n \}}=n_{1}(n_{2}-n_{1}-1)\cdots(n_{f}-n_{f-1}-1)(L-n_{f}) 
$$
with the following property
$$
\sum_{\{ n \}} D_{\{ n \}}=C_{L+1,N}\; .
$$
In order to diagonalize $A_{N}^{(\{ n \})}$ we consider an appropriate $D_{\{ n \}}$-dimensional 
basis $\{ \vert k_{1},k_{2},\cdots,k_{f+1} \rangle \}$ and write 
$$
A_{N}^{(\{ n \})} \vert \Lambda_N \rangle= \Lambda_{N}(\lambda) \vert \Lambda_N \rangle
$$
in which the eigenvectors of $A_{N}$ are written as follows
\begin{equation}
\vert \Lambda_N \rangle= \sum_{ \{k\}}
C_{k_{1},k_{2},\cdots,k_{f+1}} \vert k_{1},k_{2},\cdots,k_{f+1} \rangle \; .
\end{equation}
The coefficients can be calculated using a plane wave ansatz and one finds
$$
C_{k_{1},k_{2},\cdots,k_{f+1}}=\frac{\prod_{i=1}^{f+1} \eta^{k_{i}}\Big ( a_i(z_i)z_i^{k_{i}}+a_i(z_i^{-1}) z_i^{-k_{i}} \Big )}
{(1-\zeta)^{\delta_{k_{1},0}}(1-\xi)^{\delta_{k_{f+1},L}}}
$$
in which
$$
\begin{array}{l}
\frac{a_1(z_1)}{a_1(z_1^{-1})}=-\frac{F(z_1,\eta,\zeta)}{F(z_1^{-1},\eta,\zeta)}=-z_1^{-2n_{1}} \; , \\ \\
\frac{a_i(z_i)}{a_i(z_i^{-1})}=-z_i^{-2n_{i-1}}=-z_i^{-2n_{i}} \; \mbox{for} \; i=2,\cdots,f \; , \\ \\
\frac{a_{f+1}(z_{f+1})}{a_{f+1}(z_{f+1}^{-1})}=-z_{f+1}^{-2L}\frac{F(z_{f+1}^{-1},\eta^{-1},\xi)}{F(z_{f+1},\eta^{-1},\xi)}=-z_{f+1}^{-2n_{f}} \; .
\end{array}
$$
The eigenvalues are also given by
\begin{equation}
\label{eigenvalue AN}
\Lambda_{N}(\lambda)=(f+1)(\omega_1+\omega_2)-e^{-\lambda}\sqrt{\omega_1 \omega_2}\sum_{i=1}^{f+1}(z_i+z_i^{-1})
\end{equation}
in which $z_i$'s satisfy the following equations
\begin{equation}
\label{zs eq}
\begin{array}{l}
z_1^{2 n_{1}}= \frac{F(z_1^{-1},\eta,\zeta)}{F(z_1,\eta,\zeta)} \; , \\ \\
z_{2}^{2(n_{2}-n_{1})}=z_{3}^{2(n_{3}-n_{2})}=\cdots=z_{f}^{2(n_{f}-n_{f-1})}=1 \; , \\ \\
z_{f+1}^{2(L- n_{f})}= \frac{F(z_{f+1}^{-1},\eta^{-1},\xi)}{F(z_{f+1},\eta^{-1},\xi)}  \; .
\end{array}
\end{equation} 

In summary, the eigenvalues of $A_{N}$ can be calculated as follows: we first fix the position of the obstacles $\{ n \}$ 
which should satisfy the first relation in~(\ref{position}). We will then solve the equations~(\ref{zs eq}) and substitute their 
solutions in~(\ref{eigenvalue AN}) which gives the corresponding eigenvalues. For each set of  $\{ n \}$ one find $D_{ \{ n \} }$
eigenvalues. 

Now that all of the eigenvalues of the modified Hamiltonian~(\ref{modified hamiltonian}) are known, we discuss about the 
smallest one in the forthcoming section from which the large deviation function for the total particle current can be calculated. 

\begin{figure*}[t]
\begin{centering}
\includegraphics[width=5in]{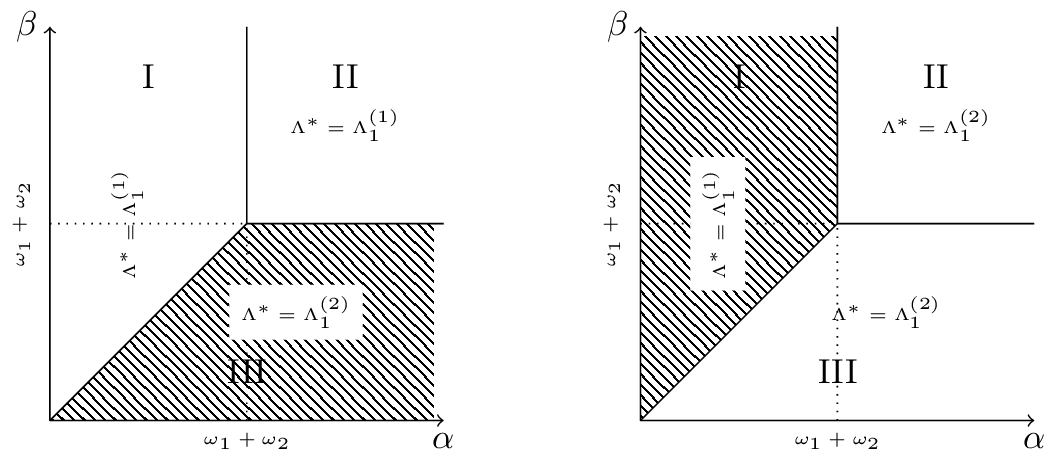}
\caption{\label{fig2} The minimum eigenvalue of the modified Hamiltonian $\tilde H$ for $\lambda > \lambda_{c}$ is given by different expressions
in different regions (see text).}
\end{centering}
\end{figure*}
%

\section{Minimum eigenvalue of $\tilde{H}$}
It is clear that the minimum eigenvalue of the modified Hamiltonian $\tilde H$ depends on both the microscopic 
reaction rates and also $\lambda$. For each value of $\lambda$ there are $2^{L}$ eigenvalues. This section is 
divided into two parts. In the first part we will consider the case $\lambda \ge 0$. The second part is devoted 
in the case $\lambda \le 0$.
\subsection{The case $\lambda \ge 0$}
The formula~(\ref{eigenvalue AN}) suggests that as $\lambda \to +\infty$ the minimum eigenvalue of $\tilde{H}$ should come 
from the eigenvalues of $A_{N}$ with the least number of random walkers, which is in this case $f=0$ i.e. the eigenvalues of 
$A_{1}$. Let us work in the thermodynamic limit $L \to \infty$. In this limit, $A_{1}$ has 
two discrete eigenvalues $\Lambda_{1}^{(1)}$ and $\Lambda_{1}^{(2)}$ which can be calculated by substituting~(\ref{z1 z2}) 
in~(\ref{eigenvalue A1}). These two eigenvalues go to zero as $\lambda \to 0$. We have found that for $\lambda \ge 0$
the minimum eigenvalue of the modified Hamiltonian is either $\Lambda_{1}^{(1)}$ or $\Lambda_{1}^{(2)}$ of $A_{1}$ depending 
on the values of the microscopic reaction rates $\omega_{1}$ and $\omega_{2}$ and also on $\lambda$. Defining 
$$
\lambda_{c} \equiv \ln[\frac{(\alpha \omega_{1}-\beta \omega_{2})^{2}+(\alpha-\beta)^{2} \omega_{1} \omega_{2}}
{(\alpha-\beta)[(\alpha \omega_{1}-\beta \omega_{2})(\omega_{1}+\omega_{2})+\alpha \beta (\omega_{2}-\omega_{1})]}]
$$
we bring a summery of the results in the following:
\begin{itemize}
\item{$\omega_1 > \omega_2$\\
In this phase, for $0 \le \lambda \le \lambda_{c}$, we have 
$\Lambda^{\ast}=\Lambda_{1}^{(1)}$. For $\lambda \ge \lambda_{c}$ the minimum eigenvalue can be determined 
using the left panel of Fig.~\ref{fig2}. As can be seen in the region III (the shaded area) the minimum eigenvalue is 
given by $\Lambda_1^{(2)}$ while it is given by $\Lambda_1^{(1)}$ in the regions I and II. The difference between 
these regions is related to the asymptotic behavior of the minimum eigenvalue $\Lambda^{\ast}$ when $\lambda \to +\infty$ which is given by
\begin{equation}
\label{asymp}
\lim_{\lambda \to +\infty}\Lambda^{\ast}(\lambda)=\left\{
\begin{array}{ll}
\alpha \quad & \mbox{in region I}, \\ 
\omega_{1}+\omega_{2} \quad & \mbox{in region II}\, ,\\
\beta \quad & \mbox{in region III} \, .
\end{array}
\right.
\end{equation} }
\item{$\omega_2 > \omega_1$\\
In this phase, for $0 \le  \lambda \le \lambda_{c}$, we have 
$\Lambda^{\ast}=\Lambda_1^{(2)}$. For $\lambda \ge \lambda_{c}$ the minimum eigenvalue can be determined 
using the right panel of Fig.~\ref{fig2}. As can be seen in the region I (the shaded area) the minimum eigenvalue is 
given by $\Lambda_1^{(1)}$ while it is given by $\Lambda_1^{(2)}$ in the regions II and III. The asymptotic behaviors 
of the minimum eigenvalue $\Lambda^{\ast}$ in different regions is given by~(\ref{asymp}).}
\end{itemize}  

Using the above description of the minimum eigenvalue $\Lambda^{\ast}$ and~(\ref{average J}) 
one can easily reproduce the results in~(\ref{average current}).

It is worth mentioning here that the first derivative of the minimum eigenvalue $\Lambda^{\ast}$ is not 
continuous at $\lambda_{c}$. This means that the large deviation function for the total particle current 
fluctuations, which can be obtained using~(\ref{GE}), is a linear function of $J$ for $J_a \le J \le J_b$~\cite{T09}
\begin{equation}
I(J)=\Lambda_{1}^{(1)}(\lambda_{c})-\lambda_{c}J=\Lambda_{1}^{(2)}(\lambda_{c})-\lambda_{c}J
\end{equation}
where for $\omega_1 > \omega_2$ 
$$J_a=\frac{d\Lambda_{1}^{(2)}}{d\lambda}\Big\vert_{\lambda=\lambda_{c}} \;\; \mbox{and} \;\;
J_b=\frac{d\Lambda_{1}^{(1)}}{d\lambda}\Big\vert_{\lambda=\lambda_{c}}\; ,$$
and that for $\omega_2 > \omega_1$  
$$J_a=\frac{d\Lambda_{1}^{(1)}}{d\lambda}\Big\vert_{\lambda=\lambda_{c}} \;\; \mbox{and} \;\;
J_b=\frac{d\Lambda_{1}^{(2)}}{d\lambda}\Big\vert_{\lambda=\lambda_{c}}.$$

The minimum eigenvalue of the modified Hamiltonian for $\lambda \ge 0$ generates, using~(\ref{GE}), the large 
deviation function for the total particle current fluctuations for $0 \le J \le \langle J \rangle$. 
\subsection{The case $\lambda \le 0$}
For $\lambda \le 0$ the situation is quite different. Let us first define $\text{Min}(\Lambda_{N})$ as the smallest 
eigenvalue of $A_{N}$. We have found that for a given finite $L$ the minimum eigenvalue of
the modified Hamiltonian $\Lambda^{\ast}$ is given by $\Lambda^{\ast}=\text{Min}(\Lambda_{2 i-1})$ for 
$\lambda_{c_{i}} \le \lambda \le \lambda_{c_{i-1}}$ in which $i=1,2,\cdots,X$ by defining
$\lambda_{c_{0}}=0$ and $\lambda_{c_{X}}=-\infty$. Here $X$ is a discrete parameter 
which maximizes
\begin{equation} 
\label{constraint}
X \cos(\frac{X \pi }{L}) \; .
\end{equation}
and that it can take one of the following values
\begin{equation}
\label{X values}
X=\left\{
\begin{array}{ll}
1,2,3,\cdots,\frac{L+2}{2} \;\; \mbox{for an even} \;\; L  \; ,   \\  \\
1,2,3,\cdots,\frac{L+1}{2} \;\; \mbox{for an odd} \;\; L  \; .
\end{array}
\right.
\end{equation}
\begin{figure*}[t]
\begin{centering}
\includegraphics[width=4in]{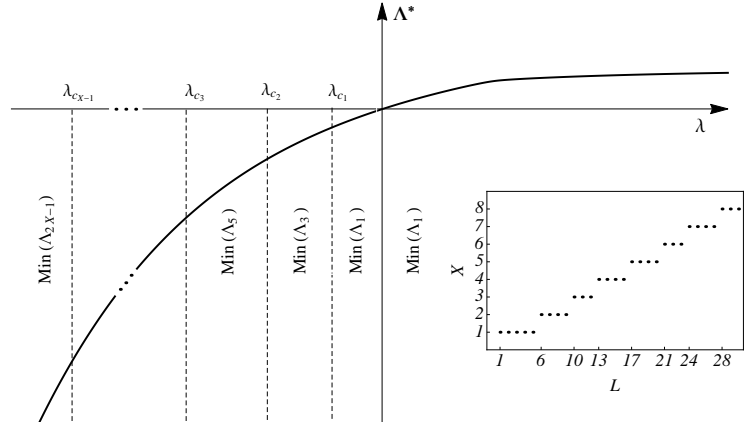}
\caption{\label{fig3}  Schematic of the structure of the minimum eigenvalue of the modified Hamiltonian $\Lambda^{\ast}$ as
a function of $\lambda$. The inset shows the maximum number of the random walkers $X$ contributing in the minimum 
eigenvalue $\Lambda^{\ast}$ for a given $L$ up to $L=30$. For more information see the text.}
\end{centering}
\end{figure*}
This means that in order to calculate the large deviation function for the total particle current in a system of length $L$, we only 
need to know the minimum eigenvalues of $A_{2i-1}$'s for $i=1,\cdots,X$. This has been shown schematically in Fig.~\ref{fig3}.
The inset of this figure shows $X$ as a function of $L$. For instance, for a system of length $21 \le L \le 23$ we only need to 
know $\text{Min}(\Lambda_{1})$ up to $\text{Min}(\Lambda_{11})$. This can be understood as follows. Let us assume that 
as $\lambda \to -\infty$ the minimum eigenvalue of the modified Hamiltonian is given by $\text{Min}(\Lambda_{2X-1})$ which
can be obtained from~(\ref{eigenvalue AN}) by substituting $N=2X-1$. $X$ is the number of random walkers which can 
be obtained from~(\ref{X values}). We are actually looking for the value of $X$ for which~(\ref{eigenvalue AN}) is minimum. 
One can easily see that as $\lambda \to -\infty$  the equations~(\ref{zs eq}) become
\begin{eqnarray*}
&& z_{1}^{2n_{1}} \simeq 1 \;, \\
&& z_{2}^{2(n_{2}-n_{1})}=\cdots=z_{X-1}^{2(n_{X-1}-n_{X-2})}=1 \;, \\
&& z_{X}^{2(L-n_{X-1})} \simeq 1
\end{eqnarray*}
in which $n_{i}$'s should satisfy~(\ref{position}). It turns out that these equations generate the minimum eigenvalue of~(\ref{eigenvalue AN}) 
provided that the distribution of the obstacles on the lattice is uniform i.e.
$$
n_{1}=n_{2}-n_{1}=\cdots=L-n_{X-1} \simeq \frac{L}{X} 
$$
which means $z_{1}=z_{2}=\cdots=z_{X}=e^{i\theta}$ with $\theta\simeq \frac{X \pi}{L}$. It is now clear that for very small and negative values of
$\lambda$,~(\ref{eigenvalue AN}) takes its minimum value provided that the expression
$$
\sum_{i=1}^{X} (z_{i}+z_{i}^{-1})=2 \sum_{i=1}^{X} \cos \theta=2 X \cos (\frac{X \pi}{L})
$$ 
becomes maximum. 
 
We have not been able to find exact analytical expressions for $\lambda_{c_{i}}$'s for $i=1,2,\cdots,X-1$; however, it 
is possible to find $\lambda_{c_{i}}$ numerically by solving the following equation
\begin{equation} 
\label{eq for lambda}
\text{Min}(\Lambda_{2i+1})\Big \vert _{\lambda=\lambda_{c_{i}}}=\text{Min}(\Lambda_{2i-1})\Big \vert _{\lambda=\lambda_{c_{i}}} \; .
\end{equation}
Our exact numerical calculations show that although the minimum eigenvalue of the modified Hamiltonian $\Lambda^{\ast}$
is continuous at $\lambda_{c_{i}}$'s, its first derivative is not continuous at these points which results in
\begin{eqnarray*}
I(J) & = & \text{Min}(\Lambda_{2i-1})\Big \vert _{\lambda=\lambda_{c_{i}}}-\lambda_{c_{i}}J \\
      & = & \text{Min}(\Lambda_{2i+1})\Big \vert _{\lambda=\lambda_{c_{i}}}-\lambda_{c_{i}}J
\end{eqnarray*}
for $J_{2i-1} \le J \le J_{2i}$ where
\begin{eqnarray*}
&& J_{2i-1} = \frac{d }{d\lambda}\text{Min}(\Lambda_{2i-1})\Big\vert_{\lambda=\lambda_{c_{i}}} \; , \\
&& J_{2i}  =  \frac{d }{d\lambda}\text{Min}(\Lambda_{2i+1})\Big\vert_{\lambda=\lambda_{c_{i}}} \; .
\end{eqnarray*}
We have also found that, for a given finite $L$, $J_{2i}-J_{2i-1}$ decreases as $i$ increases. 
This will be discussed in the next section in terms of three examples. 

\begin{figure*}[t]
\begin{centering}
\includegraphics[width=7.3in]{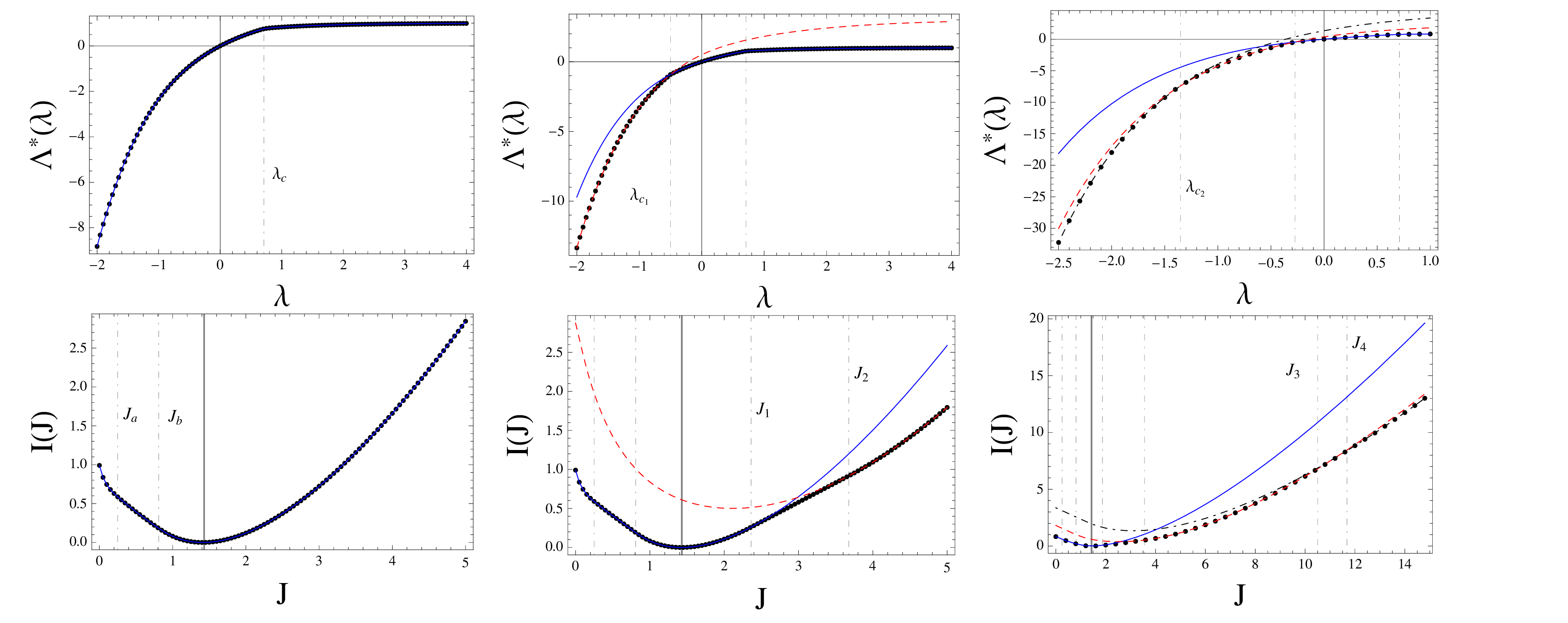}
\caption{\label{fig4} (Color online) The minimum eigenvalue of the modified Hamiltonian $\Lambda^{\ast}(\lambda)$ as a function 
of $\lambda$ (the upper row) and its corresponding large deviation function for the total particle current $I(J)$ as a function of 
$J$ (the lower row) for a system of length $L=5$, $7$ and $10$. In this figure the black dots represent the results of the numerical 
diagonalization of the modified Hamiltonian and its Legendre-Fenchel transformation. See inside the text for more details.}
\end{centering}
\end{figure*}

In what follows we will consider the large-$L$ limit which seems to be much easier to manage.  
It turns out that for $L >> 1$, $\vert \lambda_{c_{i}}-\lambda_{c_{i-1}} \vert $ for $i=2,\cdots,X-1$ drops to zero as $iL^{-2}$. 
On the other hand, it can be shown that in the thermodynamic limit $L \to \infty$ we have 
$$
\lambda^{\ast} \equiv \lambda_{c_{1}} = \lambda_{c_{2}}  = \cdots  = \lambda_{c_{X-1}} \simeq \ln 
\frac{2 \sqrt{\omega_{1}\omega_{2}}}{\omega_{1}+\omega_{2}}   \; .
$$
This means that in the large-$L$ limit, we only need to work with the minimum eigenvalues of $A_{1}$ and $A_{2X-1}$ i.e.
$$
\Lambda^{\ast}=\left\{
\begin{array}{ll}
\text{Min}(\Lambda_{1}) \;\; \mbox{for}\;\; \lambda^{\ast} \le \lambda \le +\infty \; , \\  \\
\text{Min}(\Lambda_{2X-1}) \;\; \mbox{for}\;\;  -\infty \le \lambda \le \lambda^{\ast} \; .
\end{array}
\right.
$$
At $\lambda=\lambda^{\ast}$ we have found that 
\begin{eqnarray}
\label{at large X}
\text{Min}(\Lambda_{1})\Big \vert_{\lambda=\lambda^{\ast}} & = & 
\text{Min}(\Lambda_{2X-1})\Big \vert_{\lambda=\lambda^{\ast}} \; ,\nonumber \\  
\frac{d}{d\lambda}\text{Min}(\Lambda_{1})\Big \vert_{\lambda=\lambda^{\ast}} & = &
\frac{d}{d\lambda}\text{Min}(\Lambda_{2X-1})\Big \vert_{\lambda=\lambda^{\ast}} 
\end{eqnarray}
which can be explained as follows. For $L>>1$, as $\lambda$ approaches to $\lambda^{\ast}$ from above, we 
find using~(\ref{eigenvalue A1}) that
\begin{equation}
\label{min1}
\text{Min}(\Lambda_{1})\Big \vert_{\lambda=\lambda^{\ast}}\simeq (\omega_{1}+\omega_{2})(1-\frac{1}{2}(z_{1}^{\ast}+{z_{1}^{\ast}}^{-1}))
\end{equation}
where $z_{1}^{\ast}$ is the real solution of~(\ref{eq for zs}) for $\lambda=\lambda^{\ast}$ which is given by
$G(\eta^{-1},\zeta)\vert_{\lambda=\lambda^{\ast}}$ and $G(\eta,\xi)\vert_{\lambda=\lambda^{\ast}}$ 
for $\omega_{1}>\omega_{2}$ and $\omega_{2}>\omega_{1}$ respectively. 
On the other hand, as $\lambda$ approaches to $\lambda^{\ast}$ from below we have
\begin{eqnarray}
\label{min2}
\text{Min}(\Lambda_{2X-1})\Big \vert_{\lambda=\lambda^{\ast}} &\simeq& (\omega_{1}+\omega_{2}) \Big (X \\
&& -\frac{1}{2}\sum_{i=1}^{X}(z_{i}^{\ast}+{z_{i}^{\ast}}^{-1}) \Big ) \nonumber
\end{eqnarray}
in which $z_{i}^{\ast}$'s for $i=1,\cdots,X$ are the solutions of the equations~(\ref{zs eq}) at $\lambda=\lambda^{\ast}$ 
for $f=X-1$. We have numerically checked that for $\omega_{1}>\omega_{2}$ the solutions of~(\ref{zs eq}) satisfy
$$
z_{1}^{\ast} \simeq G(\eta^{-1},\zeta)\vert_{\lambda=\lambda^{\ast}} \; , \sum_{i=2}^{X} (z_{i}^{\ast}+{z_{i}^{\ast}}^{-1}) \simeq 2(X-1)
$$
while for  $\omega_{2} > \omega_{1}$
$$
z_{X}^{\ast} \simeq G(\eta,\xi)\vert_{\lambda=\lambda^{\ast}} \; , \sum_{i=1}^{X-1} (z_{i}^{\ast}+{z_{i}^{\ast}}^{-1}) \simeq 2(X-1) \; .
$$
Replacing these into~(\ref{min2}) and comparing it with~(\ref{min1}) confirms~(\ref{at large X}).

For $\lambda << \lambda^{\ast}$ and in the large-$L$ limit, $\text{Min}(\Lambda_{2X-1})$ is approximately given by
$$
\text{Min}(\Lambda_{2X-1}) \simeq X\Big ( (\omega_{1}+\omega_{2})-2\sqrt{\omega_{1}\omega_{2}}\cos (\frac{X\pi}{L})e^{-\lambda} \Big ) \; .
$$

The minimum eigenvalue of the modified Hamiltonian for $\lambda \le 0$ generates, using~(\ref{GE}), the large 
deviation function for the total particle current fluctuations for $J \ge \langle J \rangle$. 

In the next section we will check the validity of the above mentioned results by studying three different examples.
\section{Examples: Numerical results}
In what follows we will discuss three examples in detail. These examples consist of systems of length $L=5,7$ and $10$.
Assuming $\omega_{1}=1.5$, $\omega_{2}=0.5$, $\alpha=2.5$ and $\beta=1$, the results are given in Fig.~\ref{fig4} where we have 
plotted $\Lambda^{\ast}$ and its Legendre-Fenchel transformation obtained from two different approaches. 
The first approach is the direct diagonalization of $\tilde{H}$ given by~(\ref{modified hamiltonian}) and the second 
approach is the use of~(\ref{eigenvalue AN}) by solving the equations~(\ref{zs eq}). Note that, according 
to~(\ref{average current}), the average total particle current in this phase is $\langle J \rangle \simeq 1.43$.

As can be seen in the inset of Fig.~\ref{fig3}, for a system of length $L=5$, the minimum 
eigenvalue of the modified Hamiltonian $\Lambda^{\ast}$ is given by $\text{Min}(\Lambda_{1})$ for 
$-\infty \le \lambda \le +\infty$. All other eigenvalues of $\tilde{H}$ lie above $\text{Min}(\Lambda_{1})$. 
As can be seen in the first column of Fig.~\ref{fig4} the results of the
numerical diagonalization of $\tilde{H}$ (the black dotted line) lies on the results of our analytical approach 
given by~(\ref{eigenvalue A1}) (the blue solid line). This confirms our analytical calculations for the minimum 
eigenvalue $\Lambda^{\ast}$. On the other hand, it can be seen in  Fig.~\ref{fig3} that the Legendre-Fenchel 
transformation of the minimum eigenvalue of $\tilde{H}$ obtained from these two approaches 
are exactly the same. While $\text{Min}(\Lambda_{1})$ is a continuous function of $\lambda$, its first derivative is not 
continuous at $\lambda_{c}=0.71$. As we have already explained, the large deviation function
for the total particle current, as a function of $J$, is linear where the first derivative of the minimum 
eigenvalue of the modified Hamiltonian is not continuous. The large deviation function for the total 
particle current is a linear function of $J$ for $J_{a} \le J \le J_{b}$ where $J_{a}=0.25$ and $J_{b}=0.81$.

A system of length $L=7$ is brought as the second example. As we have already explained for $\lambda \ge \lambda_{c_{1}}$
the minimum eigenvalue of the modified Hamiltonian is given by $\Lambda^{\ast}=\text{Min}(\Lambda_{1})$ while for 
$\lambda \le \lambda_{c_{1}}$ it is given by $\Lambda^{\ast}=\text{Min}(\Lambda_{3})$. Using~(\ref{eq for lambda}) we 
have obtained $\lambda_{c_{1}}=-0.50$. It can be seen in the second column of Fig.~\ref{fig4} that the minimum eigenvalue of the
modified Hamiltonian obtained from direct diagonalization of $\tilde{H}$ (the black dotted line) lies on $\text{Min}(\Lambda_{1})$ 
(the blue solid line) for $\lambda \ge \lambda_{c_{1}}$ while for $\lambda \le \lambda_{c_{1}}$ it lies on $\text{Min}(\Lambda_{3})$ 
(the red dashed line). For $\lambda \le \lambda_{c_{1}}$ the eigenvalue of $A_{3}$ given by~(\ref{eigenvalue A3}) is  the minimum 
one if we choose $n_{1}=3$ in~(\ref{z1z2 eq}). It is also clear that $\text{Min}(\Lambda_{3})$ lies below $\text{Min}(\Lambda_{1})$.
The large deviation function for the total particle current $I(J)$ is also given in the second column of Fig.~\ref{fig4}. As in the previous
example, since the minimum eigenvalue of $A_{1}$ is not differentiable at $\lambda_{c}$, the corresponding
Legendre-Fenchel transformation is a linear function of its argument for $J_{a} \le J \le J_{b}$. At $\lambda=\lambda_{c_{1}}$ the minimum
eigenvalue of $A_{1}$ is equal to the minimum eigenvalue of $A_{3}$ and, as we explained in the previous section, $\Lambda^{\ast}$
is not differentiable at this point. This results in a linear large deviation function for $J_{1} \le J \le J_{2}$ where $J_{1}=2.36$ and $J_{2}=3.67$.
The large deviation function in the second column of Fig.~\ref{fig4} has three parts. The first part for $0 \le J \le J_{1}$ (the blue solid line) comes from 
the Legendre-Fenchel transformation of $\text{Min}(\Lambda_{1})$ for $\lambda \ge \lambda_{c_{1}}$. The second part for $J_{1} \le J \le J_{2}$ 
(the black dotted line) is a linear function of $J$. Finally, the third part for $J \ge J_{2}$ (the red dashed line) comes form the 
Legendre-Fenchel transformation of $\text{Min}(\Lambda_{3})$ for $\lambda \le \lambda_{c_{1}}$. It can be seen that these 
lines lie exactly on the results obtained from Legendre-Fenchel transformation of the minimum eigenvalue of $\tilde{H}$ (the black dotted line).

Our final example is a system of length $L=10$. According to the inset of Fig.~\ref{fig3}, we need to know 
$\text{Min}(\Lambda_{1})$, $\text{Min}(\Lambda_{3})$ and $\text{Min}(\Lambda_{5})$. The minimum eigenvalue of 
$A_{1}$ for $\lambda \ge \lambda_{c_1}$ should be obtained numerically using~(\ref{eigenvalue A1}) and~(\ref{eq for zs}).
For $\lambda_{c_{2}} \le \lambda \le \lambda_{c_{1}}$ the minimum eigenvalue
of $A_{3}$ that is $\text{Min}(\Lambda_{3})$ is given by~(\ref{eigenvalue A3}) provided that we choose $n_{1}=4$ in~(\ref{z1z2 eq}). 
Finally for $\lambda \le \lambda_{c_{2}}$ the minimum eigenvalue of $A_{5}$ that is $\text{Min}(\Lambda_{5})$ 
is given by~(\ref{eigenvalue AN}) provided that we choose $n_{1}=3$ and $n_{2}=7$ in ~(\ref{zs eq}). The minimum eigenvalue of
the modified Hamiltonian obtained from direct diagonalization of $\tilde{H}$ defined in~(\ref{modified hamiltonian})
is plotted in the third column of Fig.~\ref{fig4} as a black dotted line which lies on $\text{Min}(\Lambda_{1})$ 
(the blue solid line) for $\lambda \ge \lambda_{c_{1}}$ while for $\lambda_{c_{2}} \le \lambda \le \lambda_{c_{1}}$ it lies on 
$\text{Min}(\Lambda_{3})$ (the red dashed line). For $\lambda \le \lambda_{c_{2}}$ it lies on $\text{Min}(\Lambda_{5})$
which is plotted as a black dot-dashed line. Numerical solutions of~(\ref{eq for lambda}) reveal that 
$\lambda_{c_{1}}=-0.27$ and $\lambda_{c_{2}}=-1.35$. Since $\Lambda^{\ast}$ is not a differentiable function of $\lambda$
at $\lambda_{c}$,  $\lambda_{c_{1}}$ and $\lambda_{c_{2}}$ its resulting Legendre-Fenchel transformation is a linear function of $J$ for
$J_{a} \le J \le J_{b}$, $J_{1} \le J \le J_{2}$ and $J_{3} \le J \le J_{4}$ respectively where $J_{3}=10.50$ and $J_{4}=11.68$.
As can be seen in the third column of Fig.~\ref{fig3}, the large deviation function for the total particle current obtained from 
numerically exact diagonalization of $\tilde{H}$ defined in~(\ref{modified hamiltonian}) (the black dotted line) lies exactly on the 
Legendre-Fenchel transformation of $\text{Min}(\Lambda_{1})$, $\text{Min}(\Lambda_{3})$ and $\text{Min}(\Lambda_{5})$ except where 
it has a linear behavior. One can also see that $J_{4}-J_{3} < J_{2}-J_{1}$. It turns out to be
a generic property that the distance between two consecutive $J_{i}$'s decreases as $i$ increases. 

\section{Concluding remarks}

In this paper we have considered a variant of the asymmetric zero-temperature Glauber process with open boundaries and tried 
to study the total particle current fluctuations in this system. It is known that the steady-state probability distribution vector of the system 
can be written as a linear combination of product shock measures with one shock front which performs a biased random walk on the lattice.
Using the same approach we have been able to diagonalize a modified Hamiltonian whose minimum eigenvalue generates, through a 
Legendre-Fenchel transformation, the large deviation function for the total particle current fluctuations in the system. More 
precisely, we have written the eigenvectors of the modified Hamiltonian as a linear combination of product shock measures with 
multiple shock fronts. 

Comparing our analytical results with the exact numerical calculations in three different examples, confirm the correctness of our approach. 
These examples consist of the system with three different sizes. In the first example the minimum
eigenvalue of the modified Hamiltonian can be obtained by diagonalizing it in an invariant sector which is constructed by 
the product shock measures with a single moving shock front. In order to calculate this minimum eigenvalue in the second example
we need to diagonalize the modified Hamiltonian in another invariant sectors which is constructed by
the product shock measures with one and two moving shock fronts. In the third example the minimum eigenvalue of the modified 
Hamiltonian should be obtained by diagonalizing it in an invariant sectors consists of the product shock measures with one,
two and three moving shock fronts.

As we mentioned, the system studied in this paper is a special variant of the asymmetric zero-temperature Glauber process 
introduced in~\cite{KJS}. The reactions in this process consist of only forward particle hopping. It would be interesting to 
investigate the particle current fluctuations in the full process where backward particle hopping is also included.

\appendix
\section{\label{A1} Matrix elements of $A_N$ and $B_N$}

The matrix elements of $A_{N}$ and $B_{N}$ can be obtained using the following relations for an odd $i$
\begin{widetext}
$$
\begin{array}{l}
\langle k'_{1},n'_{1},...,n'_{i-1},k'_{i},n'_{i},...,n'_{f},k'_{f+1} \vert A_{N} \vert k_{1},n_{1},...,n_{i-1},k_{i},n_{i},...,n_{f},k_{f+1}\rangle = \\
\Big(\frac{N+1}{2}(\omega_{1}+\omega_{2})\Big)^{(1-\delta_{k_1,0})(1-\delta_{k_{f+1},L})}
\Big(\alpha+\beta+\frac{N-3}{2}(\omega_{1}+\omega_{2})\Big)^{\delta_{k_1,0}\delta_{k_{f+1},L}(1-\delta_{N,1})}\\
\Big(\alpha+\frac{N-1}{2}(\omega_{1}+\omega_{2})\Big)^{\delta_{k_1,0}(1-\delta_{k_{f+1},L})}
\Big(\beta+\frac{N-1}{2}(\omega_{1}+\omega_{2})\Big)^{\delta_{k_{f+1},L}(1-\delta_{k_1,0})} 
\prod_{j=1}^{f+1} \delta_{k'_j,k_j} \prod_{r=1}^{f} \delta_{n'_r,n_r}\\
-\Big((\omega_{1}e^{-\lambda})^{1-\delta_{k_{f+1},L}\delta_{i,f+1}}\beta^{\delta_{k_{f+1},L}\delta_{i,f+1}}\delta_{k'_{i},k_{i}-1}
+(\omega_{2}e^{-\lambda})^{1-\delta_{k_1,0}\delta_{i,1}}\alpha^{\delta_{k_1,0}\delta_{i,1}}\delta_{k'_{i},k_{i}+1} \Big)
\prod_{j=1,j\ne i}^{f+1} \delta_{k'_j,k_j} \prod_{r=1}^{f} \delta_{n'_r,n_r} \; , \\ \\ \\
\langle k'_{1},n'_{1},...,n'_{f-1},k'_{f}\vert B_{N}\vert k_{1},n_{1},...,n_{i-1},k_{i},n_{i},...,n_{f},k_{f+1}\rangle =\\
-\Big(
(\omega_{1}e^{-\lambda})^{1-\delta_{k_{f+1},L}\delta_{i,f+1}}\beta^{\delta_{k_{f+1},L}\delta_{i,f+1}}\delta_{n_{i-1},k_{i}-1}
\prod_{j=1}^{i-1} \delta_{k_j,k'_j}\prod_{j=i+1}^{f+1} \delta_{k_j,k'_{j-1}} 
\prod_{r=1}^{i-2} \delta_{n_r,n'_r}\prod_{r=i}^{f} \delta_{n_r,n'_{r-1}} \\
+ (\omega_{2}e^{-\lambda})^{1-\delta_{k_1,0}\delta_{i,1}}\alpha^{\delta_{k_1,0}\delta_{i,1}}\delta_{n_{i},k_{i}+1}
\prod_{j=1}^{i-1} \delta_{k_j,k'_j}\prod_{j=i+1}^{f+1} \delta_{k_j,k'_{j-1}}
\prod_{r=1}^{i-1} \delta_{n_r,n'_r}\prod_{r=i+1}^{f} \delta_{n_r,n'_{r-1}}
\Big) \; .
\end{array}
$$
\end{widetext}
These matrix elements are zero for an even $i$.

\section{\label{A2} Matrices $\tilde{A}_{1}$ and $\tilde{A}_{2}$}
The matrix representations for $\tilde{A}_{1}$ and $\tilde{A}_{2}$ are 
\begin{widetext}
$$
\begin{array}{l}
\tilde{A}_{1}=\left( \begin{array}{ccccccc}
\alpha+\omega_1+\omega_2&-\omega_1 e^{-\lambda}&0&\cdots&0&0&0\\
-\omega_2 e^{-\lambda} & \alpha+\omega_1+\omega_2&-\omega_1 e^{-\lambda}&\cdots&0&0&0\\
0&-\omega_2 e^{-\lambda} & \alpha+\omega_1+\omega_2&\cdots&0&0&0\\
\vdots&\vdots&\vdots&\ddots&\vdots&\vdots&\vdots\\
0&0&0&\cdots&\alpha+\omega_1+\omega_2&-\omega_1 e^{-\lambda}&0 \\
0&0&0&\cdots&-\omega_2 e^{-\lambda}&\alpha+\omega_1+\omega_2&-\beta \\
0&0&0&\cdots&0&-\omega_2 e^{-\lambda}&\alpha+\beta
\end{array} \right)\, \; , \\  \\
\tilde{A}_{2}=\left( \begin{array}{ccccccc}
2(\omega_1+\omega_2)&-\omega_1 e^{-\lambda}&0&\cdots&0&0&0\\
-\omega_2 e^{-\lambda} &2(\omega_1+\omega_2)&-\omega_1 e^{-\lambda}&\cdots&0&0&0\\
0&-\omega_2 e^{-\lambda} &2(\omega_1+\omega_2)&\cdots&0&0&0\\
\vdots&\vdots&\vdots&\ddots&\vdots&\vdots&\vdots\\
0&0&0&\cdots&2(\omega_1+\omega_2)&-\omega_1 e^{-\lambda}&0 \\
0&0&0&\cdots&-\omega_2 e^{-\lambda}&2(\omega_1+\omega_2)&-\beta \\
0&0&0&\cdots&0&-\omega_2 e^{-\lambda}&2(\omega_1+\omega_2)
\end{array} \right)\,   \;.
\end{array}
$$
\end{widetext}

\section*{Acknowledgment} 
S. R. M. would like to thank Islamic Azad University Hamadan for their financial support. 

\end{document}